\documentclass[12pt]{article}
\usepackage{graphicx}
\begin{document}
\title{Density Profile and Flow of Miscible Fluid with Dissimilar Constituent Masses}

\maketitle
\centerline{\bf R.B. Pandey$^{1,2}$, D. Stauffer$^2$, R. Seyfarth$^2$,}
\centerline{\bf Luis A. Cueva$^2$, J.F. Gettrust$^1$, Warren Wood$^1$}

\bigskip
\centerline{$^1$Naval Research Laboratory}
\centerline{Stennis Space Center, MS 39529}
\bigskip
\centerline{$^2$Department of Physics and Astronomy}
\centerline{University of Southern Mississippi,
Hattiesburg, MS 39406-5046}
\maketitle
\begin{abstract}
A computer simulation model is used to study the
density profile and flow of a miscible gaseous fluid mixture consisting
of differing constituent masses ($m_A = m_B/3$) through an open matrix.
The density profile is found to decay with the height
$\propto \exp(-m_{A(B)}h)$, consistent with the barometric height law.
The flux density shows a power-law increase $\propto
{(p_c-p)}^{\mu}$ with $\mu \simeq 2.3$ at the porosity $1-p$ above the
pore percolation threshold $1-p_c$. 

\end{abstract}
\newpage
\section{Introduction}

Understanding the flow of a complex gaseous mixtures, 
sedimentation, and evolution of density profiles of its 
constituents in geo-marine systems and near-surface 
ecological environments is becoming increasingly important
\cite{1,2}. There are a number of examples:
($i$) High density brines associated with salt tectonics in large salt 
provinces (e.g. the Gulf of Mexico) have been breaching the seafloor 
and forming pools toxic to native flora and fauna \cite{macdonald}.
This involves flow of a mixture of fluids with different densities due 
to different salt content and temperature. ($ii$) A mixture of air and 
radon flux 
through unsaturated soils within the upper few meters of the land 
surface.  Radon 222, one of the intermediate products of the decay of 
uranium 238 to lead 206, is an odorless, radioactive gas (with half 
life 3.8 days), and is common in many soils and rocks.  Because radon 
is about 8 times more dense than air, and is relatively inert 
\cite{crc}, it easily penetrates porous building materials in ground 
floors and basements, especially when pressure gradients are created 
by central heating systems \cite{epa}. The U.S. Environmental 
Protection Agency (EPA) estimated in 1986 that 5,000 to 20,000 persons 
in the United States die of lung cancer each year from inhaling 
radioactive radon decay products in homes and buildings \cite{epa}. 
($iii$) Evidence of methane hydrate formation below the ocean 
floor and in sub-ocean bottom in mud-volcano involves the 
flow and sedimentation of complex gas and fluid mixtures with 
dissimilar masses \cite{methane, gins, vogt}. Studies of flow and 
density distribution of a mixture of miscible gas and fluids through 
porous media are therefore highly desirable.

Systematic studies based on the field measurements of flow and density 
profile of gas and fluid constituents in geomarine environment are 
severely limited \cite{methane} due to uncontrollable changes in 
geothermal parameters and morphological variations. Thus, computer 
simulation studies remain one of the most viable tools to probe such 
difficult issues as flow \cite{stanley, sahimi, redner} and 
density profile \cite{pbg}. Incorporating all details 
(multiple constituents and their characteristics) even with a coarse 
grained host matrix i.e. open porous media with appropriate 
concentration gradient and pore distributions \cite{methane} is a 
challenging issue.  
Lattice gas and particulate methods in general (Boltzmann, Cellular 
Automata, Ising (interacting), etc.) \cite{holian, rothman} have been 
used in diverse applications of fluid flow. In study of the density 
profile and fluid flow of an interacting fluid mixture through porous 
media, a direct application of traditional hydrodynamics approaches
\cite{3,4} becomes 
intractable; apart from a major problem of enormous boundary 
conditions in such a porous medium (percolating system), it is not clear how to include interaction or reaction between the fluid components in 
hydrodynamic equations. Interacting lattice gas \cite{pbg} may be a 
simple approach to initiate probing such difficult issues.

Very recently we studied the flow of a fluid described by particles, 
say of type "A" through an open porous medium using a computer 
simulation model \cite{previous}. The porous medium is generated
by a random distribution of sediment particles of concentration $p$ on a 
simple cubic lattice. The bottom layer of the matrix is connected 
to a source of mobile fluid particles ("A"). As soon as a bottom site
becomes empty, it is immediately filled by a particle "A" from the 
source.  Particles in the bottom layer are not allowed to move below 
this plane due to presence of abundant source particles. On the other 
hand, the particles
can escape the system from the top if they attempt to move to the
higher layer. In this concentration driven system, the flux density 
shows a power-law decay with the porosity near percolation threshold. 
The steady-state density profile of fluid particles depends systematically on
the barrier concentration $p$.

In this article, we extend our previous studies \cite{previous} to a miscible 
two-component
system consisting of constituents, say "A" and "B" with dissimilar masses ($m_a$
and $m_b, \; m_b = 3 m_a$).  The model is described in the following section 2.
We incorporate the effect of gravitational 
potential energy in moving the particles and allowing them to escape the
system from the bottom layer as well. 
The injection probability of particles $A$ and $B$ at the bottom remains equal.
The results are presented in section 3 with a conclusion in section 4.

\section{Model}
Each site on a simple-cubic $L \times L \times L$ lattice, with $L$ up to 250,
can be in one of four states:
occupied by an A particle, occupied by a B particle, empty (0), or a barrier 
site. Nearest-neighbor particles interact with energy $J$ such that A and B mix
well: $J(A,A) = J(B,B) = -J(A,B)= -J(B,A) = -J(A,0) = -J(B,0) = 1$ where 
negative $J$ means
attraction and positive $J$ means repulsion. The immobile barriers exert no 
force on the particles except to prevent them from occupying the barrier site.
The energy thus is
$$ E = \sum_i \sum_k J(I,K)$$
where $i$ runs over all particles, $k$ over all neighbor sites of $i$, and $I$ 
and $K$ are the corresponding site variables (A, B, barrier, or 0).

\begin{figure}[hbt]
\begin{center}
\includegraphics[angle=-90,scale=0.35]{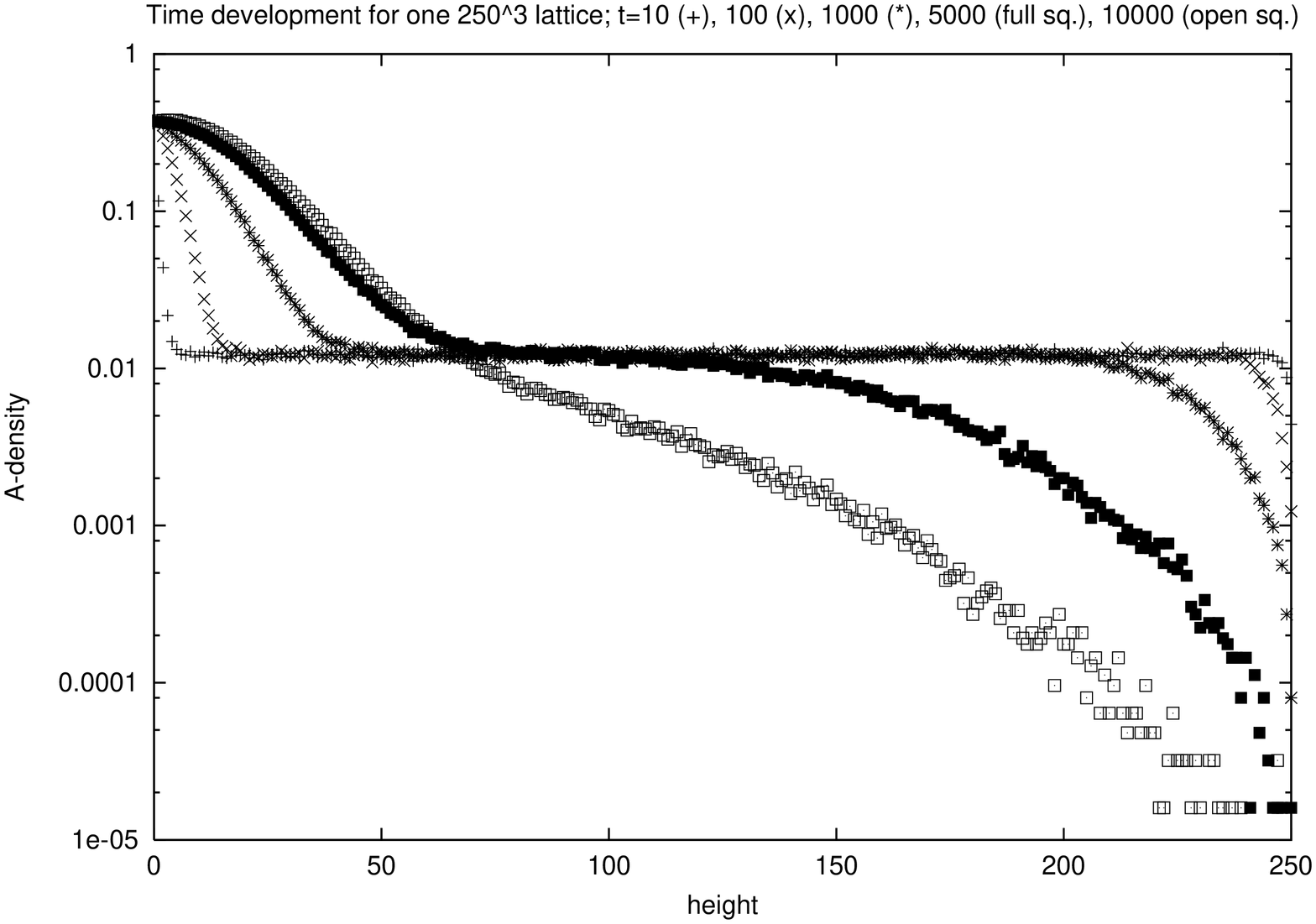}
\includegraphics[angle=-90,scale=0.35]{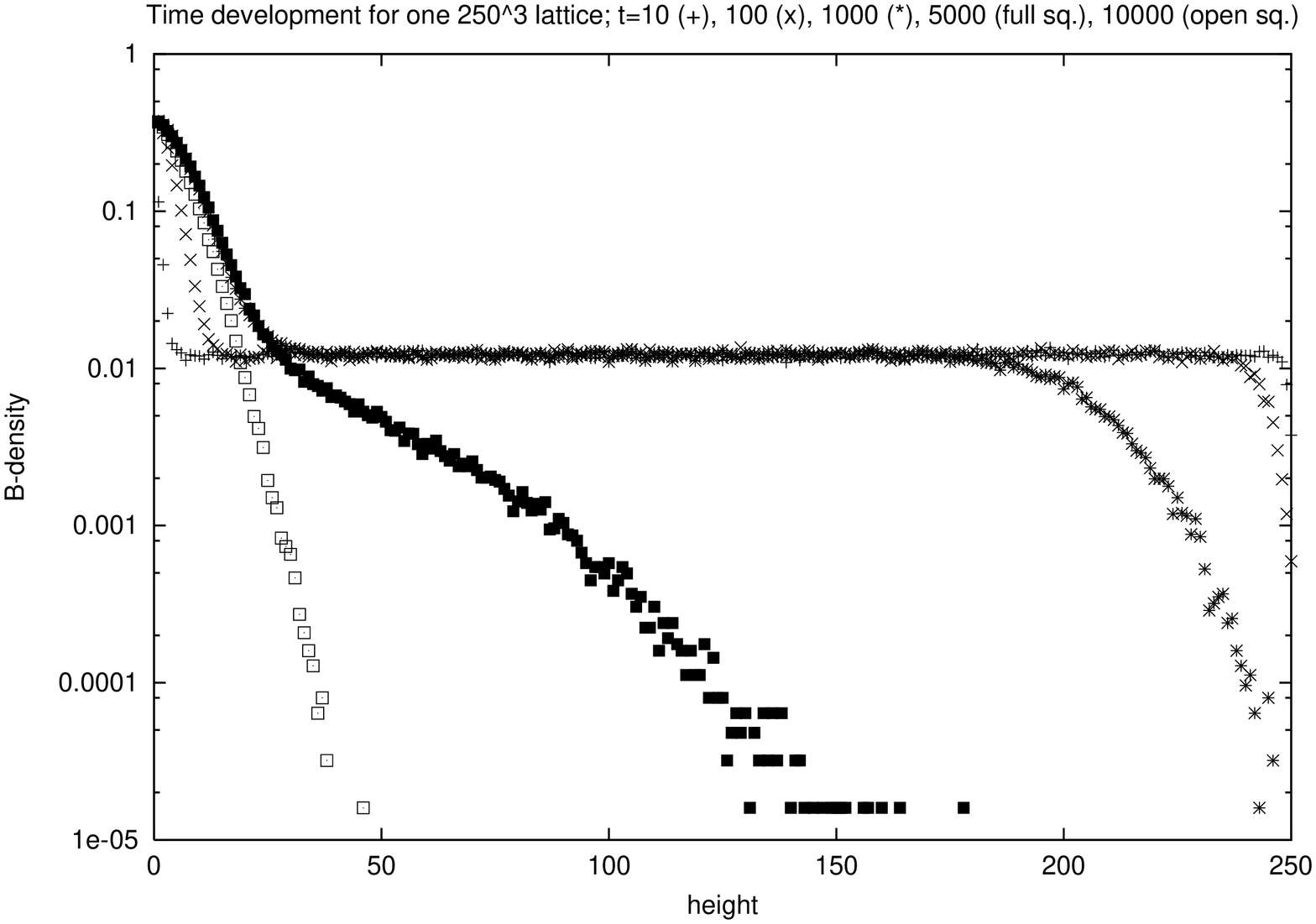}
\end{center}
\caption{
Crossover from constant to exponential density profiles, with 
increasing times $t$ as given in headline, for A-particles (part a) 
and B-particles (part b). No sites are barriers in this simulation. 
}
\end{figure}

A and B particles can move with a Metropolis probability exp$(-\Delta E/k_BT)$ 
to neighboring sites; here $\Delta E$ is the energy change associated with
this move and $k_BT = 5$ is the thermal energy.  If a particle on the lowest 
plane moves upwards or horizontally, we inject a new particle A or B 
(with equal probability) onto the vacated site.
If a particle on the lowest plane wants to move downward, it drops out. 
A particle moving upward from the highest plane (relatively a rare event)
is lost, without
any injection from the top. Periodic boundary conditions are applied in the
horizontal directions. In this way, an $ \infty \times \infty \times L$ plate is
approximated, with new material injected from the bottom at a nearly constant
rate.

Gravity pulls down the particles through an energy $mk_BT$ where the lattice 
constant and the gravitational constant are incorporated into the dimensionless
mass $m=0.1$ for A and $m=0.3$ for B particles. Thus the barometric height
law gives an equilibrium density $\propto \exp(-mh)$ as a function of height
$h, \; 1 < h < L$.

One time step is an attempt to move each particle once (on average) through 
random sequential updating; it does not matter much if instead we 
enforce exactly one attempt per particle for each time step. For $L=30$ we 
used up to $t=10^6$ time steps, without seeing any long-time effects; for 
larger $L$ (50 to 250) typically $t = 10^4$ to $10^5$ gave equilibrium. 
Initially, the lattice is occupied homogeneously with a low concentration of 
particles, half A and half B. 
Our computer program allows many more choices for interactions and boundary 
effects and is developed to investigate many different systems (details are
available from ras.pandey@usm.edu).
One diffusion attempt, without barrier sites, took about 0.5 microseconds 
on one Cray-T3E processor.

\section{Results}

\begin{figure}[hbt]
\begin{center}
\includegraphics[angle=-90,scale=0.32]{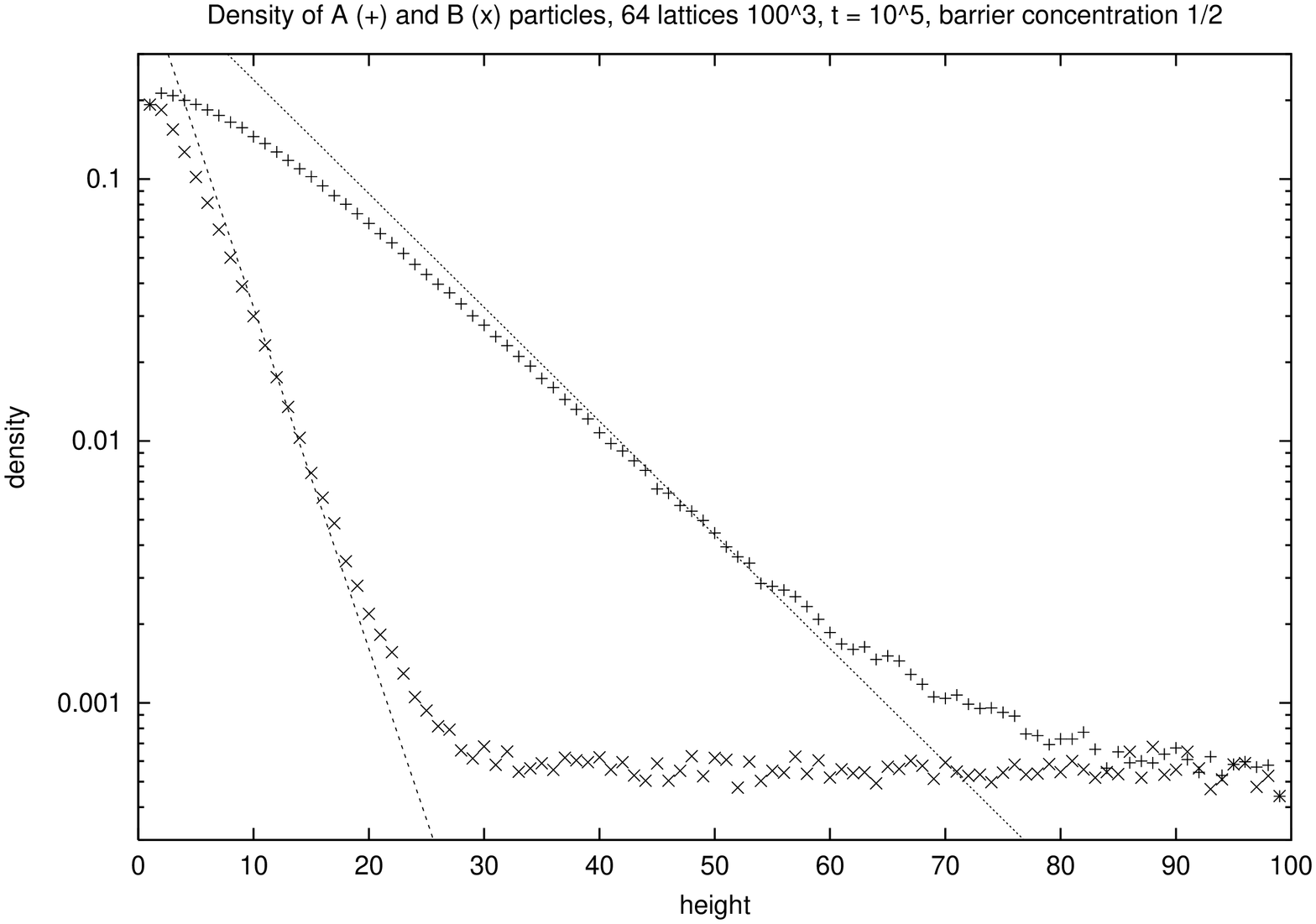}
\includegraphics[angle=-90,scale=0.32]{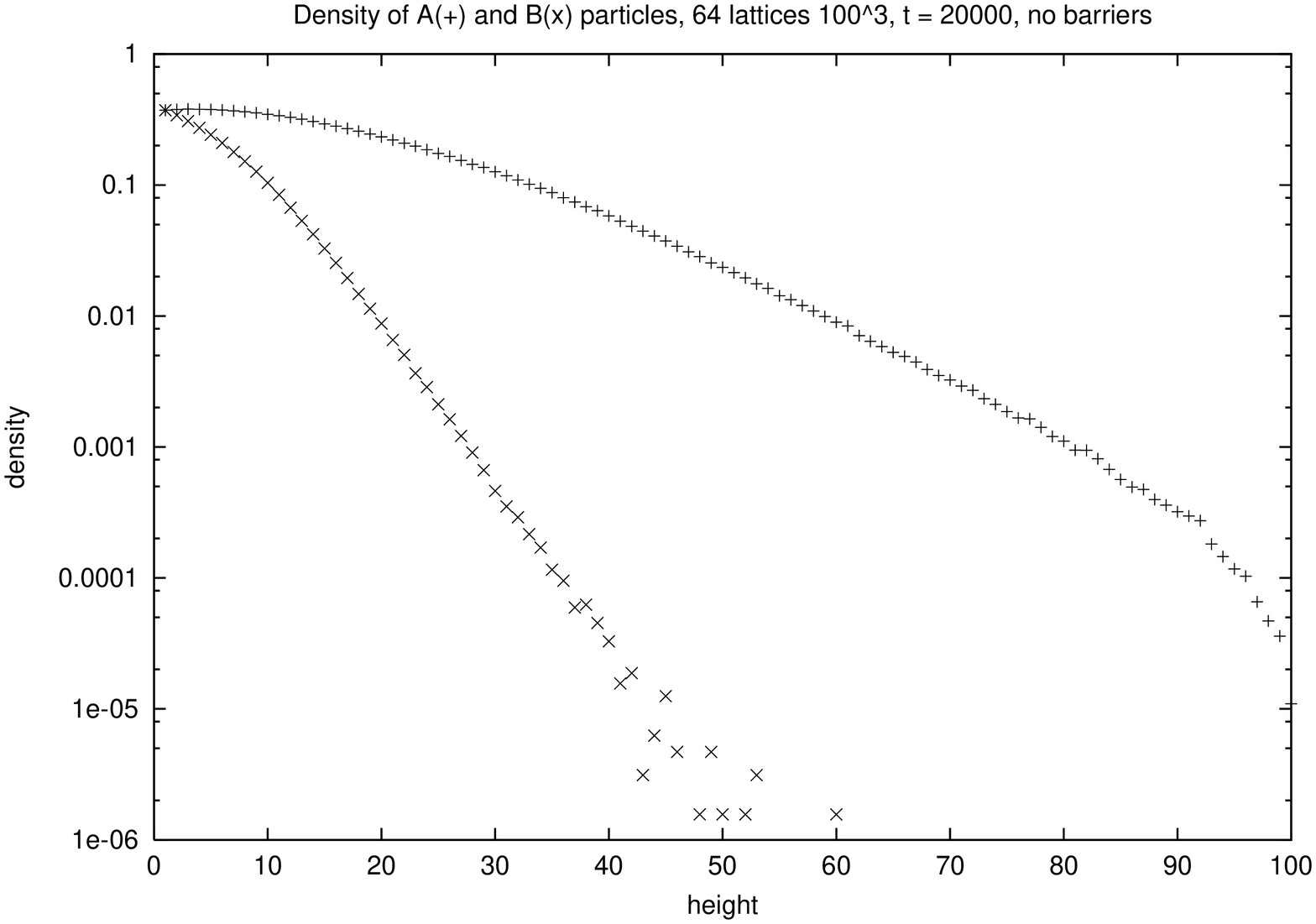}
\end{center}
\caption{
Equilibrium density profiles with (part a) and without (part b) half the
sites occupied by barriers. The straight lines are $\propto \exp(-mh)$. 
}
\end{figure}

Fig.1 shows for our largest lattices how the initial constant density profiles
change with time, starting from the two boundaries, into a roughly exponential 
decay for intermediate heights. The equilibrium density profiles, Fig.2,
at intermediate densities are consistent with the barometric height law shown
as straight lines in these semilogarithmic plots, both with and without barrier
sites. The nearest-neighbor correlation functions (not shown), i.e. the number
of A or B particles surrounding a particle at height $h$, decay qualitatively 
similar to the density profiles.

\begin{figure}[hbt]
\begin{center}
\includegraphics[angle=-90,scale=0.5]{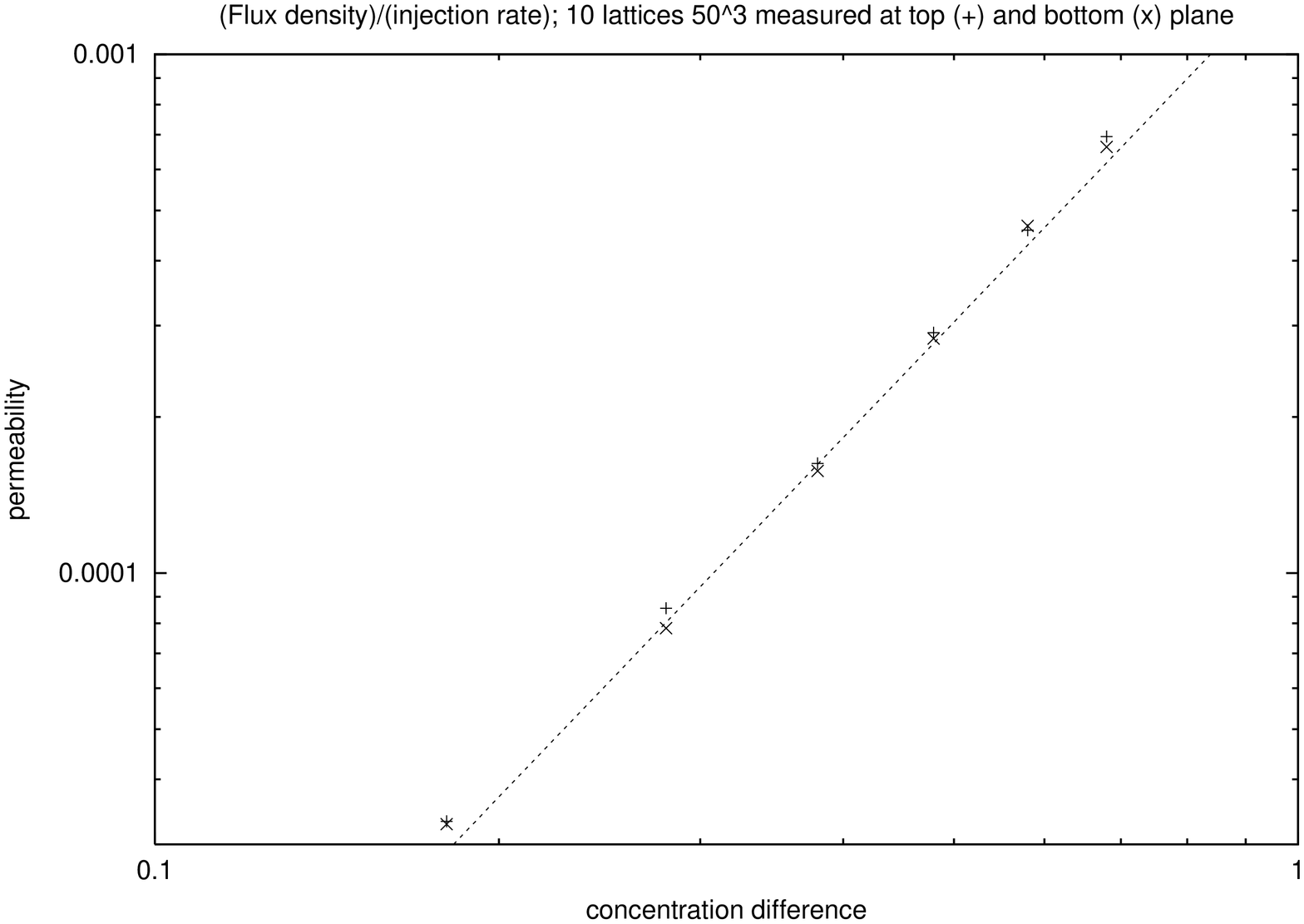}
\end{center}
\caption{
Double-logarithmic plot of permeability versus $p_c-p$ where $p_c = 
0.6884$ is the percolation threshold and $p$ the barrier concentration. For 
$p >p_c$ there is no infinite connected set of fluid sites between the 
barriers and thus the permeability vanishes. The straight line has a slope 2.3.
}
\end{figure}

\begin{figure}[hbt]
\begin{center}
\includegraphics[angle=-90,scale=0.5]{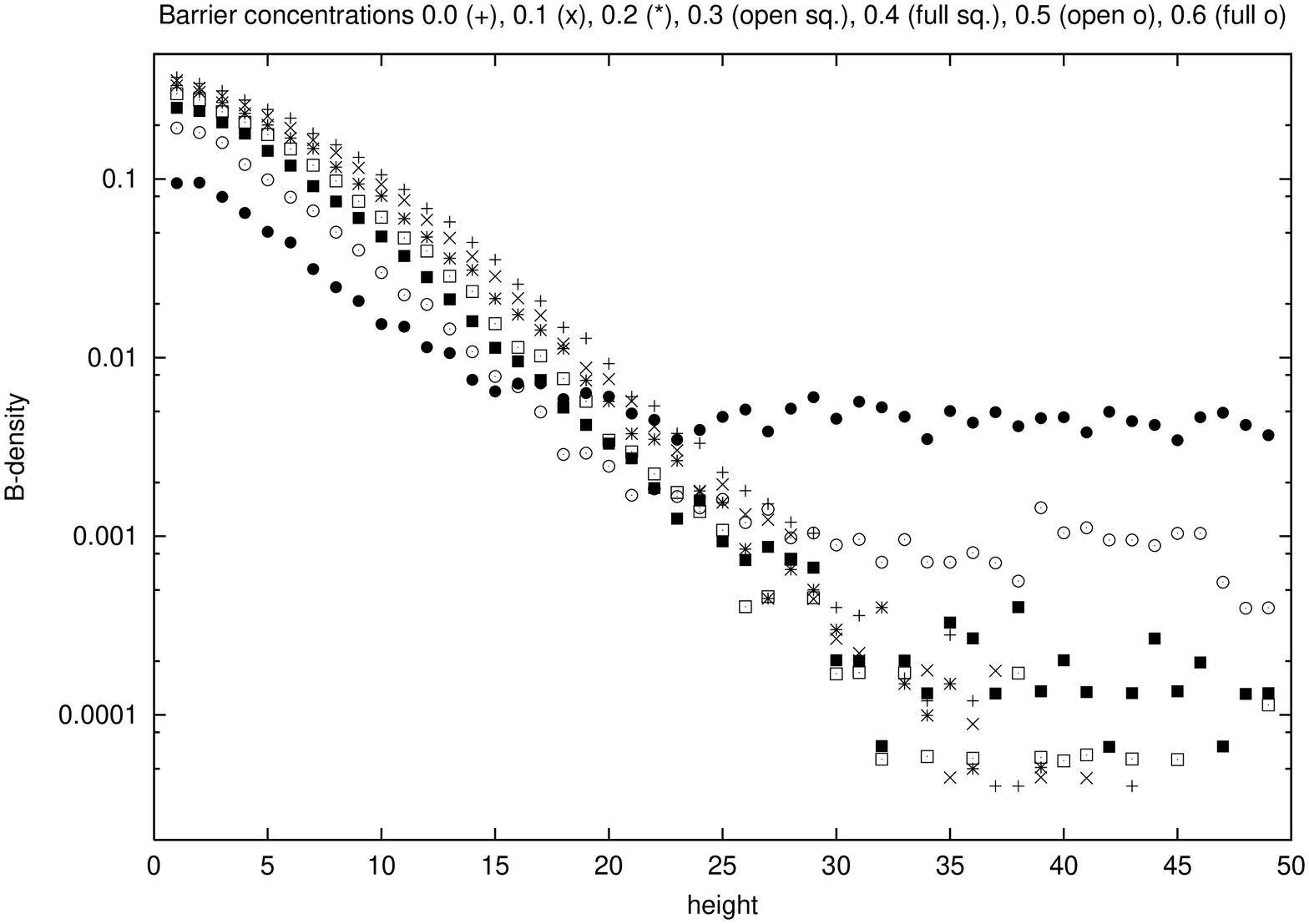}
\end{center}
\caption{
Equilibrium density profiles of B-particles for various barrier
concentrations (same simulations as for Fig.3). 
}
\end{figure}

Fig.3 shows as a function of the barrier concentration $p$ the system's 
permeability, defined as the net flow (per unit time and unit cross-sectional
area) at the top or bottom surfaces divided by the injection rate at the 
lowest plane.  The straight line in this log-log plot suggests a power law 
similar to that of the electrical conductivity in random resistor networks.
Fig.4 shows the density profiles of the B particles
for the same simulations. We see there an exponential decay, followed by a 
plateau whose value strongly increases with increasing $p$, suggesting
a trapping of particles by the barrier sites. We have also looked at the
velocity and instantaneous velocity profiles of the two constituents which
are consistent with our expectations, i.e., more mobility toward the top.

\section{Conclusion}

A computer simulation model is used to study the flow rate, sedimentation, 
and density profile of a mixture of miscible particle systems for a range of 
porosities above the pore percolation threshold in an open porous medium.
In our concentration driven system, the steady-state density profiles for
both $A$ and $B$ fluid are reached. Both density profiles show the well-known
exponential decay $\propto \exp(-m_{A(B)}h)$ with height $h$. The effect of mass
difference is vividly illustrated in the density profiles: while for our
lattice sizes the density
of $A$ particles continues to decay up to the top plane the density of $B$
particles already saturates at some low value. 
The saturated density of B particles increases on decreasing 
the porosity $1-p$ - we speculate this saturation is due to
trapping of particles in the pores.

The flux of particles $A$ at the bottom becomes equal to the outward flux from 
the top in steady state.  The flux density of particles $A$ decays with 
porosity with a power-law $\propto (p_c - p)^\mu$ with $\mu \simeq 2$.

\bigskip
{\it Acknowledgments}: 

We acknowledge partial supports from ONR PE\# 0602435N and DOE-EPSCoR grant.
This work was supported in part by  grants of computer
time from the DOD High Performance Computing Modernization
Program at the Major Shared Resource Center (MSRC), NAVO,
Stennis Space Center, and from the Julich supercomputer center.


\begin{thebibliography}{99}
\bibitem{1} 
R. Sassen, S. Joye, S.T. Sweet, D.A. DeFreitas, A.V. Molkov, I.R.
MacDonald, Organic GeoChem. 30, 485 (1999).

\bibitem{2} 
H.H. Roberts and R. Carney, Economic Geology 92, 863 (1997).

\bibitem{macdonald} MacDonald, I.R., Reilly J.F., II, Guinasso, N.L., Jr., 
Brooks, J.M., Carney, R.S., Bryant, W.A. and Bright, T.J., Science 248, 1096 
(1990)
\bibitem{crc} Chemical Rubber Co. Handbook of Chemistry and Physics, CRC press;
www.hbcpnetbase.com
\bibitem{epa} Environmental Protection Agency (EPA), 1986, 
A citizen's guide to radon: 14 p.; http://www.epa.gov/iaq/radon/pubs/citguide.html.
\bibitem{methane} {\em ``Natural Gas Hydrates''}, Geophysical Monograph 124,
edited by C.K. Paull and W.P. Dillon, American Geopysical Union (2001). 
\bibitem{gins} G.D. Ginsburg, A.D. Milkov, V.A. Soloviev, A.V. Egorov,
G.A. Cherkashev, R.P Vogt, K. Crane, T.D. Lorenson, and M.D. Khutorskoy,
Geo-Marine Letters 19, 57 (1999).
\bibitem{vogt} P.R. Vogt, J. Gardner, and K. Crane, Geo-Marine Letters 
19, 2 (1999).
\bibitem{stanley} H.E. Stanley and J.S. Andrade Jr., Physica A 295, 17 (2001).

\bibitem{sahimi}M. Sahimi, {\em ``Flow and transport in Porous Media and
 Fractured Rock''} (VCH Weinheim, 1995). 

\bibitem{redner} J. Koplik, S. Redner, and D. Wilkinson, Phys. Rev. A 
37, 2619 (1988).

\bibitem{pbg} R.B. Pandey, J.L. Becklehimer, and J.F. Gettrust,
Physica A 289, 321 (2001).

\bibitem{holian}
{\em "Microscopic Simulations of Complex Hydrodynamic Phenomena"}, eds.
     M. Mareschal and B. Holian, Plenum, New York (1992).

\bibitem{rothman}
D. H. Rothman and S. Zaleski, {\em ``Lattice-Gas Cellular Automata -- 
Simple Models of Complex Hydrodynamics''} (Cambridge University Press, 1997).
\bibitem{3}
E.L. Cussler, {\em ``Diffusion: Mass transfer in fluid systems''}
(Cambridge University Press, 1984).

\bibitem{4}
C.A. Silebi and W.E. Schiesser, {\em ``Dynamic Modeling of Transport
Process Systems''} (Academic Press, Inc., 1992).

\bibitem{previous} R.B. Pandey, J. F. Gettrust, and D. Stauffer, Physica
A 300, 1 (2001)
\end{thebibliography}
\end{document}